\documentclass[review,number,sort&compress]{elsarticle}
\usepackage{lineno,hyperref}
\usepackage{csquotes}
\usepackage{placeins}
\usepackage{afterpage}
\usepackage{graphics}
\usepackage{color}
\usepackage{subfigure}
\usepackage{textgreek}

\begin{document}
\begin{frontmatter}

\title{Low energy recoil detection with a spherical proportional counter}


\author[a]{I.Savvidis}
\author[a,b]{I.Katsioulas\corref{mycorrespondingauthor}}
\cortext[mycorrespondingauthor]{Corresponding author}
\ead{ioannis.katsioulas@cea.fr}
\author[a]{C.Eleftheriadis}

\author[b]{I.Giomataris}
\author[b]{T.Papaevangellou}
\address[a]{Aristotle University of Thessaloniki, Greece }
\address[b]{IRFU, CEA, Universit\'e Paris-Saclay, F-91191 Gif-sur-Yvette, France   }
%




\begin{abstract}
We present results for the detection of low energy nuclear recoils in the keV energy region, from measurements performed with the Spherical Proportional Counter (SPC). An ${}^{241}$Am-${}^{9}$Be fast neutron source is used in order to obtain neutron-nucleus elastic scattering events inside the gaseous volume of the detector. The detector performance in the keV energy region was measured by observing the 5.9 keV line of a ${}^{55}$Fe X-ray source, with energy resolution of $10\%$ ($\sigma$). 
The toolkit GEANT4 was used to simulate the irradiation of the detector by an ${}^{241}$Am-${}^{9}$Be source, while SRIM was used to calculate the Ionization Quenching Factor (IQF), the simulation results are compared with the measurements. The potential of the SPC in low energy recoil detection makes the detector a good candidate for a wide range of applications, including Supernova or reactor neutrino detection and Dark Matter (WIMP) searches (via coherent elastic scattering).
\end{abstract}


\end{frontmatter}


\section{Introduction}
The detection of very low energy nuclear recoils is essential to the field of direct neutrino detection and of direct dark matter searches. During the last years the dark matter hunt is continued with increasing rate. Weakly Interacting Massive Particles (WIMPs) are an eminent candidate for Dark Matter (DM). A common way of search for this type of particles relies on the elastic scattering of WIMPs on the target nuclei. The recoil energy is  expected to be in the keV (or less) energy range, for WIMPs with a mass lower than 10 GeV. Another important field of searches concomitant to that of DM searches \citep{gerbier1} (being also  a background source to them), is the direct detection of low energy neutrinos (1 MeV - 100 MeV). Supernova neutrinos, solar, geoneutrinos and reactor neutrinos all belong to this category. The Standard Model neutrino-nucleon interaction was proposed years ago and it is gaining in popularity lately because of the large cross section it provides through the coherence effect, where all nucleons contribute to the scattering (especially the neutrons), resulting in a cross section increased by the neutron number squared. Again the detection of these neutrinos relies on the observation of the recoiling nuclei in the keV energy range and depending on the neutrino source, observing recoils of a few hundred eV (reactor neutrinos for example \citep{vergados1,kosmas1}). These applications require a detector with a very low detection energy threshold ($\sim100$ eV). We propose the utilization of the SPC, a spherical gaseous detector recently developed by Giomataris et al \citep{giomataris1}, for applications with such requirements. In this work, we present the capability of the SPC in the detection of low energy nuclear recoils (keV - 150 keV energy region), as well as the simulated detector response to low energy recoils, taking into account the effect of the Ionization Quenching Factor (IQF) \citep{hitachi} by using GEANT4 \citep{agostinelli} and SRIM \citep{srim}. 

\begin{figure}[!h]
\centering
\includegraphics[width=60mm]{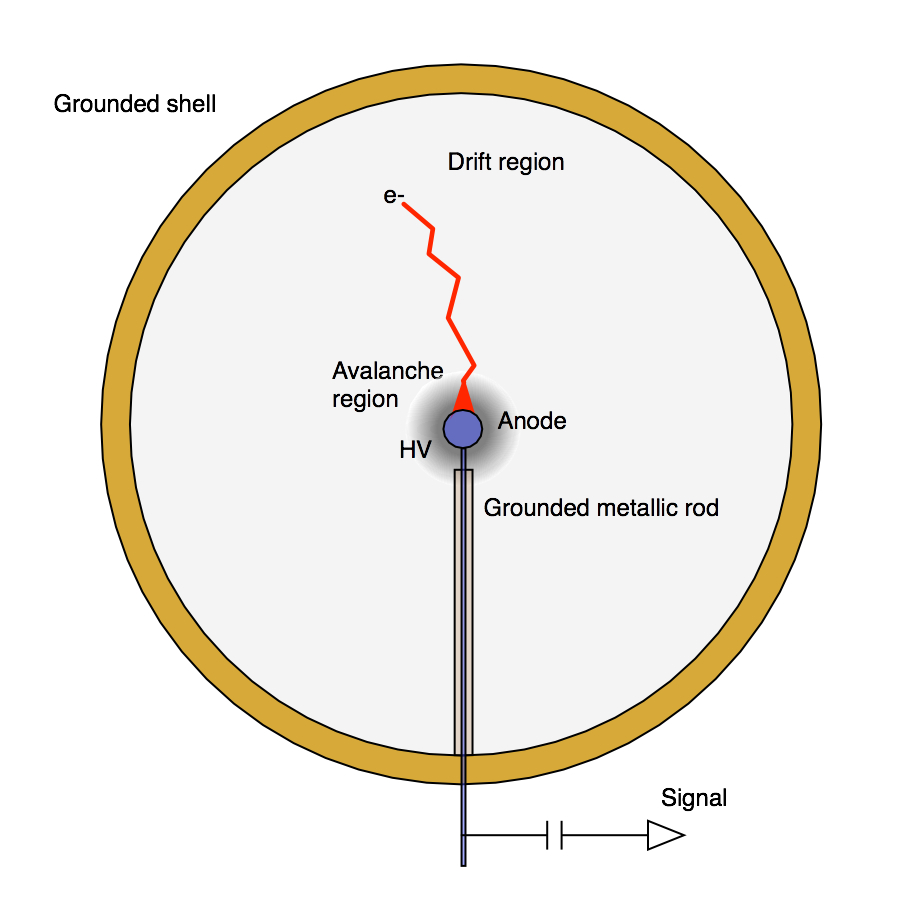}
\caption{A schematic of the Spherical Proportional Counter and the detection principle.}
\label{spc}
\end{figure}

\section{The spherical proportional counter}

The detector consists of a spherical vessel which is grounded and filled with a gas mixture (up to 5 bar pressure). The anode which consists of a small ball (usually made from a metallic or a resistive material) is placed in the center of the vessel and supported by a grounded metallic rod, through which the high voltage is applied. Thus the electric field is varying with the reverse of the distance squared ($1/r^{2}$) and it is highly inhomogeneous along the radius. The difference in the intensity of the electric field, from the outer to the inner radii, divides the detector volume into two regions, the drift and the amplification region (figure \ref{spc}). Primary ionization electrons in the drift region, drift towards the anode (the drift time varies from \textmu s to ms depending on the gas mixture and pressure). When these electrons reach a distance of a few mm from the anode, the avalanche starts due to the intense electric field. The pulse shape of the signal depends on the charge spatial density distribution and the distance of the interaction from the anode. 
The main advantages of this detector, for low energy recoil detection (either neutrino or WIMP induced recoils), are the simple design, the large volumes achievable, the energy resolution in the low energy region ($<10\%$ at 5.9 keV) \citep{bougamont1,bougamont2} and the low electronic noise provided by the low capacitance due to the spherical geometry \citep{andriamonje} (even for large detector sizes). The low energy threshold of the detector is limited only by the mean ionization energy of the gas mixture  \citep{bougamont1}. The detection range varies from a few eV to tenths of MeV, depending on the amplification field, allowing detection of low energy gammas, electrons, alpha particles and heavy ions. Lastly, the fiducialization capability of the detector \citep{bougamont3} allows us to distinguish point like energy depositions (low energy gammas, low energy electrons, heavy ions) from spatially extended depositions (muons) through pulse shape analysis. A complete description of the SPC (principle of detection, characteristics and capabilities) can be found at \citep{giomataris1}.

\section{Experimental setup and low energy calibration}

\begin{figure}[h]
\centering
\includegraphics[width=60mm]{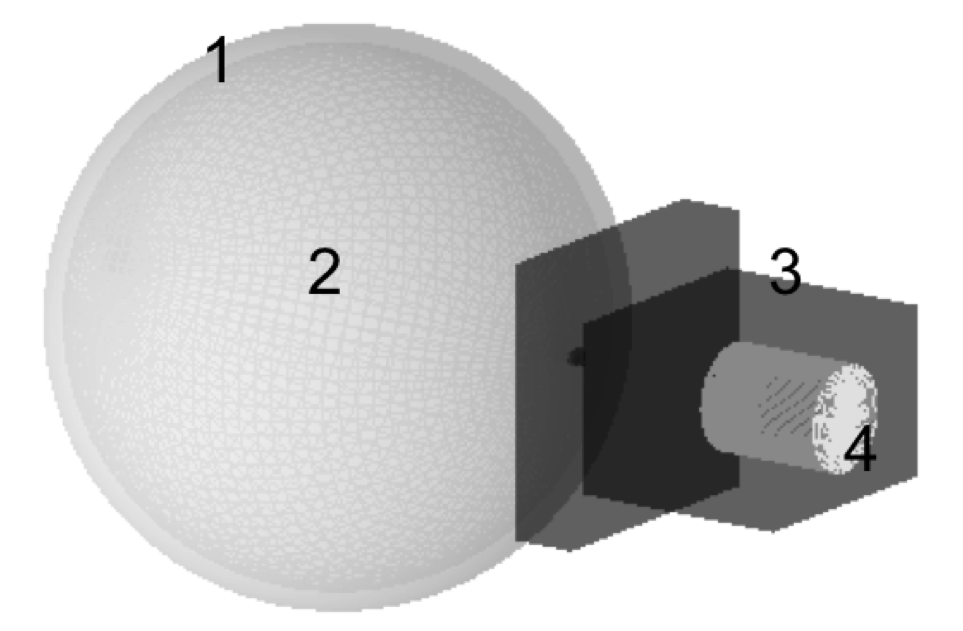}
\caption{A schematic of the detector setup: 1) The detector vessel, 2) the gaseous volume, 3) the lead shielding and 4) the source case.}
\label{schema}
\end{figure}

A SPC placed at the Aristotle University of Thessaloniki, was used for this study. The spherical vessel of the detector is 40 cm in diameter and made from 1.5 cm thick Duran 2.3 glass (coated with a graphite layer for electric conduction). The anode in the center of the cavity is metallic and 2 mm in diameter. The electronics used during these measurements were the CANBERRA 2006 charge sensitive preamplifier (50 \textmu s fall time) and an Amplitude to Digital Converter used to readout the output of the preamplifier and to digitize the pulses, which were then registered to a computer memory. The vessel was filled with Ar:CH${}_{4}\ (98:2)$ gas mixture at pressure up to 500\ mbar. The low energy calibration of the detector was performed using a ${}^{55}$Fe source (5.9 keV), the fluorescence lines of ${}^{27}$Al (1.45 keV) \citep{bougamont2} and ${}^{241}$Am (13.95 keV, 17.7 keV) \citep{liu}. To test the response of the detectors to neutrinos or WIMPs, nuclear recoils have to be produced, for this reason the detectors were exposed to an ${}^{241}$Am-${}^{9}$Be source with an activity of $5.94\times10^4$ neutrons per second. The ${}^{241}$Am-${}^{9}$Be source also emits gamma rays, the most important line being the 4.44 MeV line from the deexcitation of the ${}^{12}$C* nucleus produced from the Be(\textalpha ,n) reaction \citep{mowlavi}. The intensity of this line is directly related to the neutron intensity with a ratio $R=S_{g}/S_{n} = 0.591\pm\ 2.6\%$ \citep{croft}. The source was placed 15 cm away from the surface of the detector. To prevent gamma ray contamination, it was cased inside a lead castle of 12 cm thickness along the axis (figure \ref{schema}). 

\FloatBarrier
\section{Results $-$ Pulse shape analysis}
\begin{figure}[h]
\centering
\includegraphics[width=90mm]{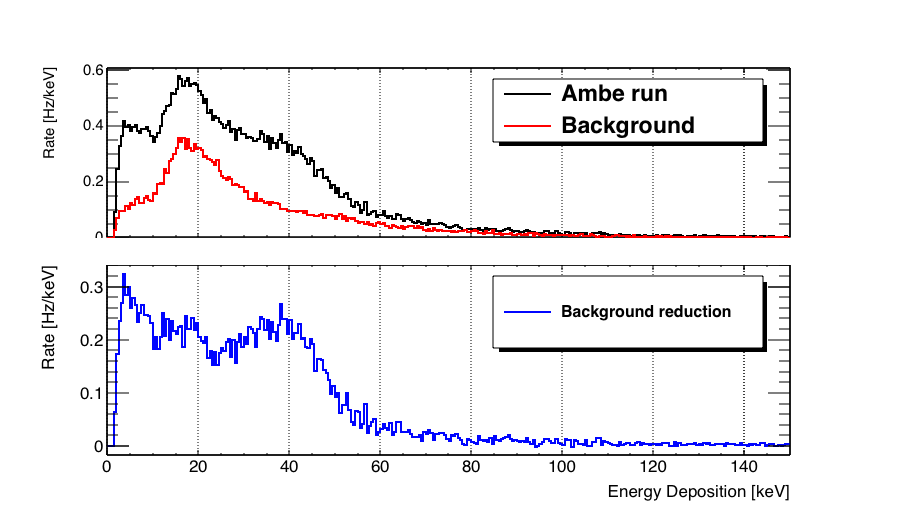}

\caption{Top diagram: energy deposition spectra of the ${}^{241}$Am-${}^{9}$Be run (black line) and the background (red line). Bottom:  difference between the two. No cuts are applied.}
\label{edepNocuts}
\end{figure}

\begin{figure}[h]
\centering
\includegraphics[width=90mm]{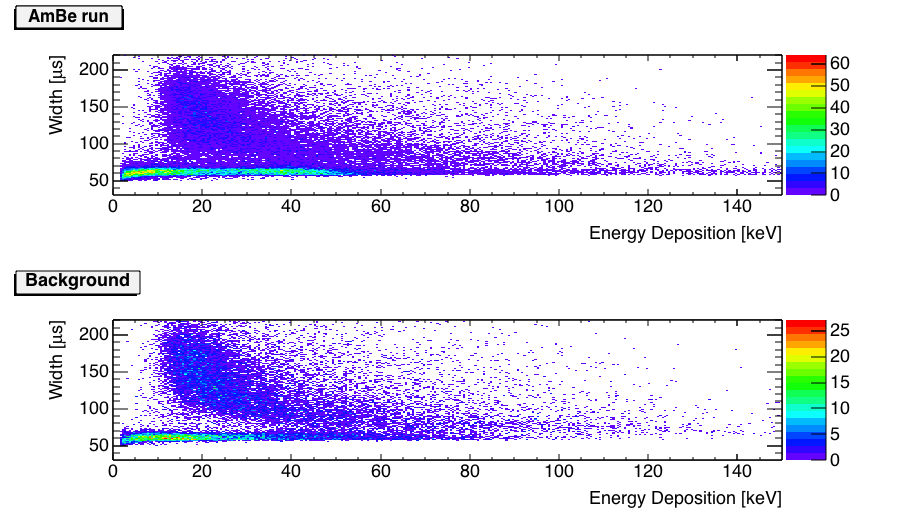}
\caption{The width of the pulse versus energy deposition for the ${}^{241}$Am-${}^{9}$Be run (top) and the background run (bottom), without any cuts.}
\label{edepWidNocuts}
\end{figure}

\begin{figure}[]
\centering
\includegraphics[width=90mm]{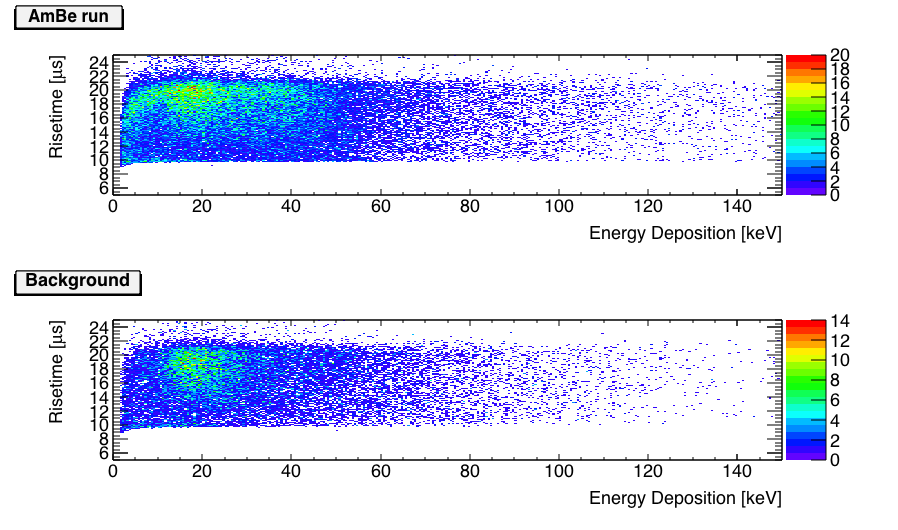}
\caption{The pulse rise time versus energy deposition for the ${}^{241}$Am-${}^{9}$Be run (top) and the background run (bottom), without any cuts.}
\label{edepRtNocuts}
\end{figure}

To attain the results presented below, the detector was filled with Ar:CH${}_{4}\ (98:2)$ at 500 mbar and was irradiated by the ${}^{241}$Am-${}^{9}$Be source (for a period of 3600 s); due to cosmic radiation (ie atmospheric muons \citep{grieder} and neutrons) and natural radioactivity from the surrounding walls (ie ${}^{40}$K, ${}^{238}$U and ${}^{232}$Th daughter isotopes) the background contribution had to be measured (also for a period of 3600 s), in order to be subtracted from the recorded signal. The data acquired during the measurements were analyzed using Pulse Shape Analysis (PSA). The parameters used in the analysis were a) the pulse height, which is used to estimate the energy deposition of an event, b) the pulse rise time, which is the time interval between $10\%$ and $90\%$ of pulse height, c) the Full Width at Half Maximum (FWHM) of the pulse and d) the number of "peaks" or local maxima in a pulse, which is calculated from the number of zero crossings of the pulse derivative. The pulse rise time and width correspond to the dispersion of the primary electron drift time (time interval between their production and their arrival to the anode under the influence of the electric field).  Figures \ref{edepNocuts}, \ref{edepWidNocuts} and \ref{edepRtNocuts} show the results of the analysis without any pulse shape cuts. The data acquisition threshold set corresponds to an $\sim2$ keV energy detection threshold. The atmospheric muon contribution to the energy deposition spectra is visible at figure \ref{edepNocuts} (top diagram) as a "peak" with energy around 20 keV.     
 
\begin{figure}[h]
  \centering
  \subfigure[]{\includegraphics[width=80mm]{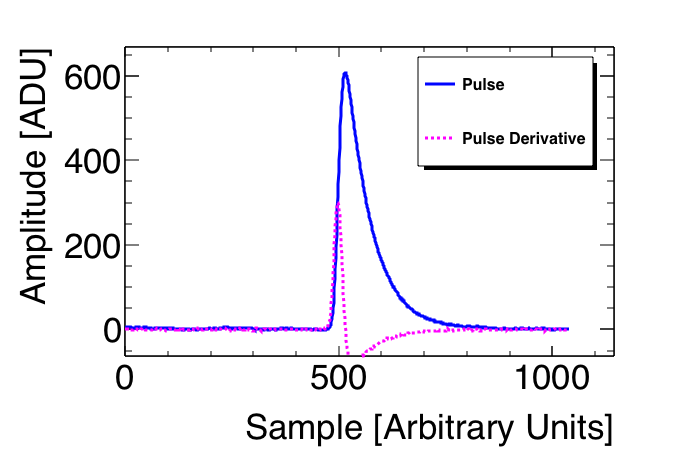}\label{mu_pulsea}}
  \\
  \subfigure[]{\includegraphics[width=80mm]{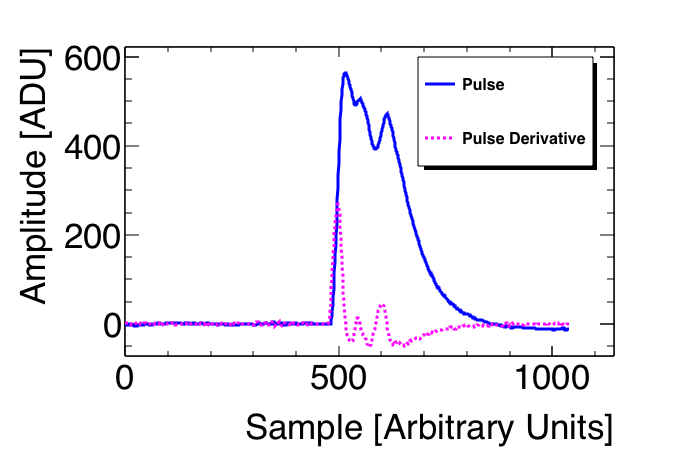}\label{mu_pulseb}}
  \caption{Measured pulses corresponding to an event a) with point like energy deposition and b) with an energy deposition along a spatially extended track, displaying multiple peaks.}
  \label{mu_pulse}
\end{figure}

Particles with enough energy to cross the whole length of the detector such as atmospheric muons and energetic electrons deposit energy all along their lengthy track ($\sim$ tenths of cm). They can also produce delta rays energetic enough to ionize further at a distance from the primary track, creating ionization clusters with large charge density. This kind of "behavior" is translated to pulses with higher width than punctual energy depositions (figure \ref{mu_pulsea}), which may contain multiple peaks (due to variations in drift time and in spatial ionization density), as for example the pulse presented in figure \ref{mu_pulseb}. 


\begin{figure}[!h]
\centering
\includegraphics[width=80mm]{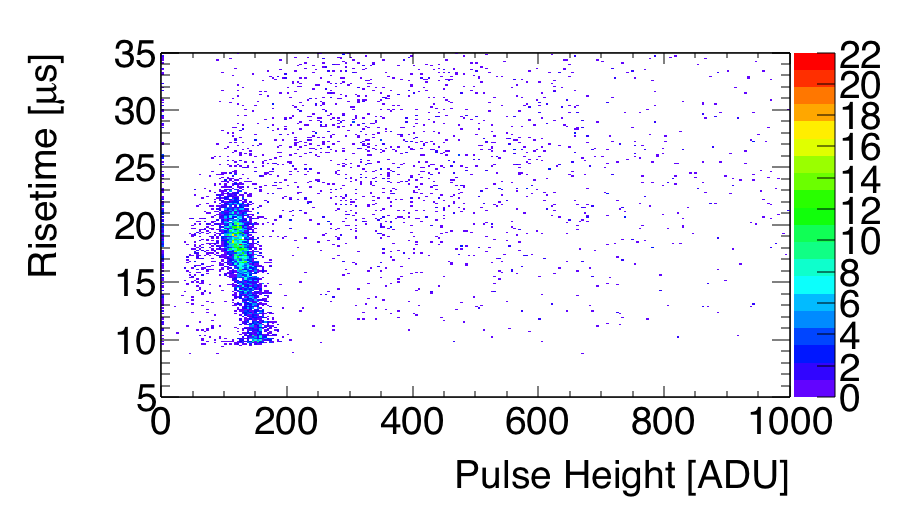}
\caption{The pulse rise time versus the pulse height of the ${}^{55}$Fe calibration run.}
\label{calib1}
\end{figure}

\begin{figure}[!h]
\centering
\includegraphics[width=80mm]{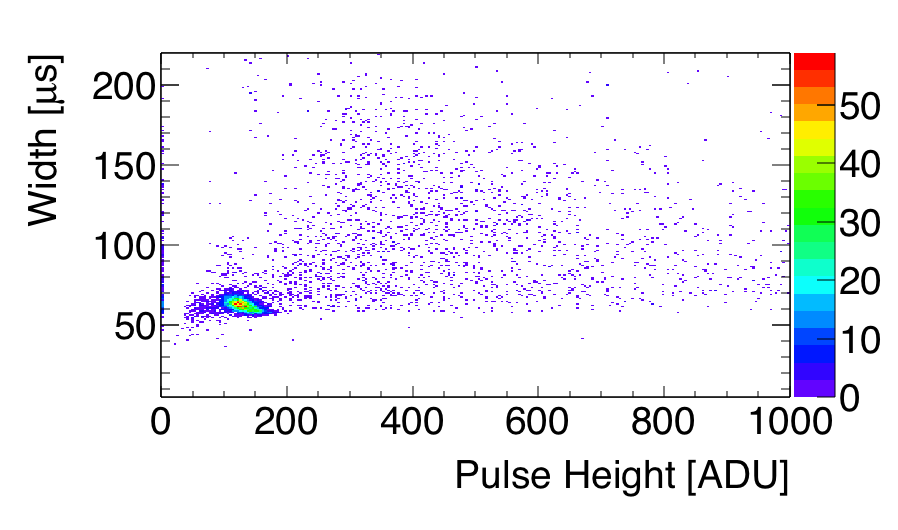}
\caption{The pulse width versus the pulse height of the ${}^{55}$Fe calibration run.}
\label{calib2}
\end{figure}

\begin{figure}[!h]
\centering
\includegraphics[width=80mm]{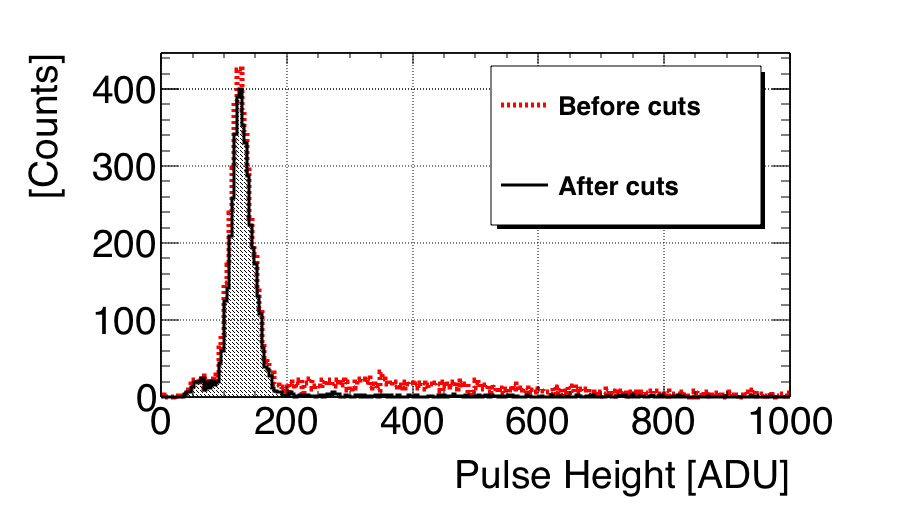}
\caption{The pulse height spectrum of the ${}^{55}$Fe calibration run a) before applying any pulse shape cuts (red) and b) after applying the pulse shape cuts (black).}
\label{calib3}
\end{figure}

\begin{figure}[h]
\centering
\includegraphics[width=90mm]{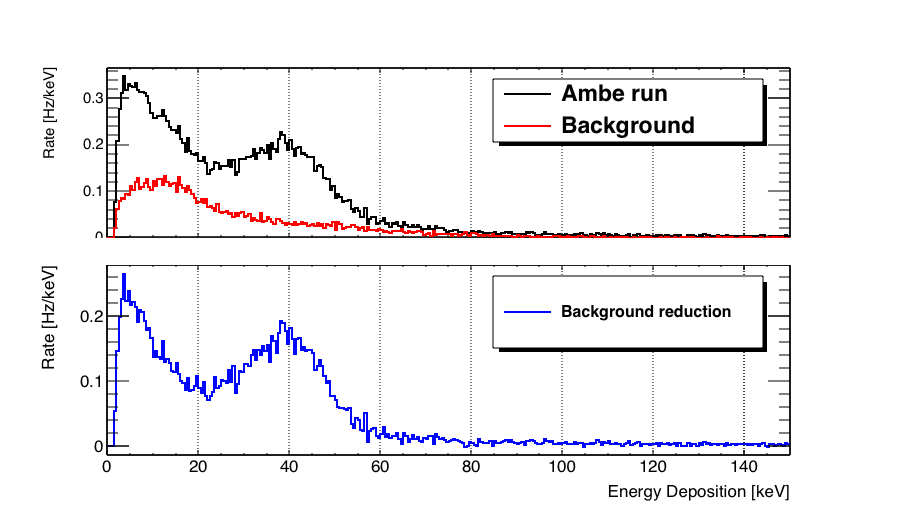}
\caption{Top diagram: energy deposition spectra of the ${}^{241}$Am-${}^{9}$Be run (black) and the background (red). Bottom:  difference between the two (blue). The results are shown after the pulse shape cuts.}
\label{edepcuts}
\end{figure}

\begin{figure}[h]
\centering
\includegraphics[width=90mm]{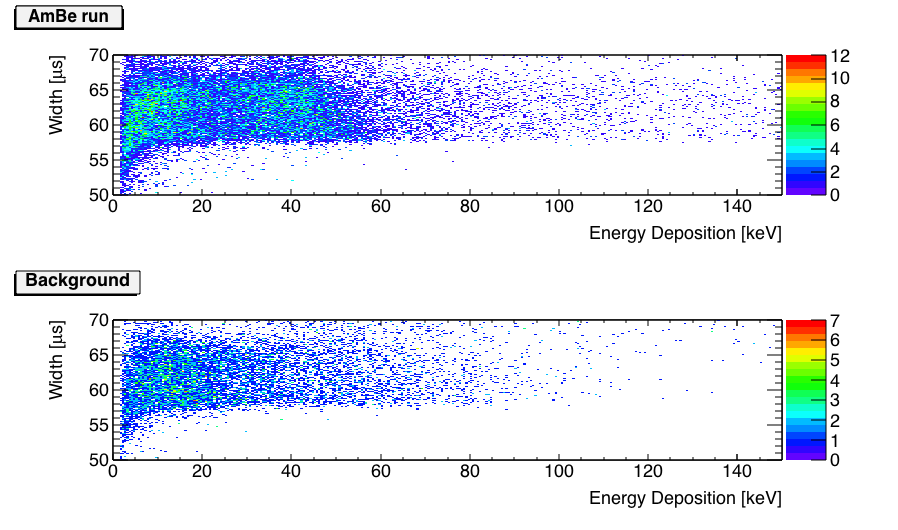}
\caption{The width of the pulse versus energy deposition of the pulse histograms for the ${}^{241}$Am-${}^{9}$Be run (top) and the background (bottom), after applying the pulse shape cuts.}
\label{edepWidcuts}
\end{figure}

\begin{figure}[h]
\centering
\includegraphics[width=90mm]{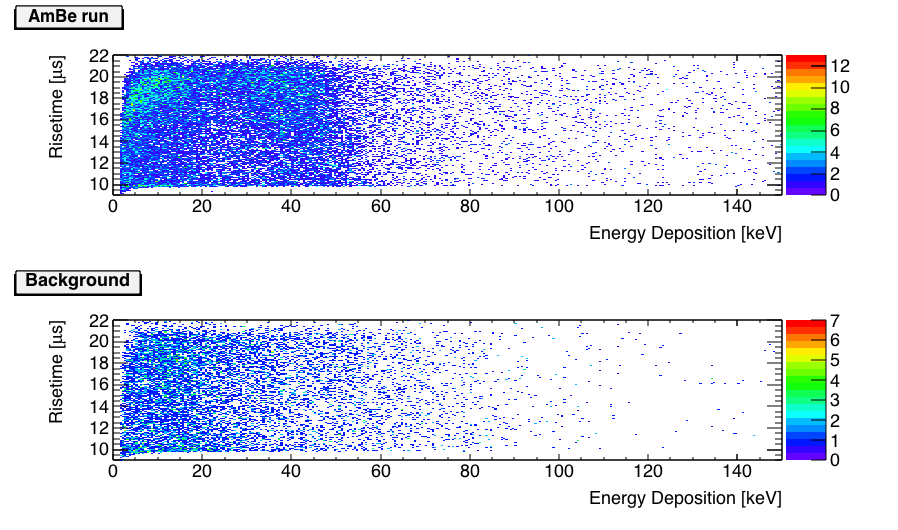}
\caption{The pulse rise time versus energy deposition for the ${}^{241}$Am-${}^{9}$Be run (top) and the background (bottom), after applying the pulse shape cuts.}
\label{edepRtcuts}
\end{figure}

\FloatBarrier
The muon contribution can be minimized by rejecting events outside specific pulse width and pulse rise time intervals and also rejecting events with multiple peaks. These rejection intervals can be determined by looking at the calibration measurements, as for example the width versus energy deposition and rise time versus energy deposition plots of figures \ref{calib1}, \ref{calib2}  which correspond to the calibration measurements with a 5.9 keV ${}^{55}$Fe X-ray source. Utilizing the information provided by the analysis of the calibration measurements, one can infer that events with width outside the 50 \textmu s and 70 \textmu s interval,  pulse rise time outside the 11 \textmu s and 21 \textmu s interval and with multiples peaks should be rejected. A comparison between the pulse height spectra before and after the pulse shape cuts is presented in figure \ref{calib3}. Events corresponding to extended energy depositions (compared to the short ranged events of the 5.9 keV electron events of the ${}^{55}$Fe X-ray line) are rejected while retaining more than $85\%$ of the measured X-ray line. The results after performing the pulse shape cuts to ${}^{241}$Am-${}^{9}$Be run data and the background run data are presented in figures \ref{edepcuts}, \ref{edepWidcuts} and \ref{edepRtcuts}.

\FloatBarrier
\section{Simulations $-$ Comparison with experimental results}
\label{simul}
\subsection{The GEANT4 simulation}
\begin{figure}[h]
\centering
\includegraphics[width=90mm]{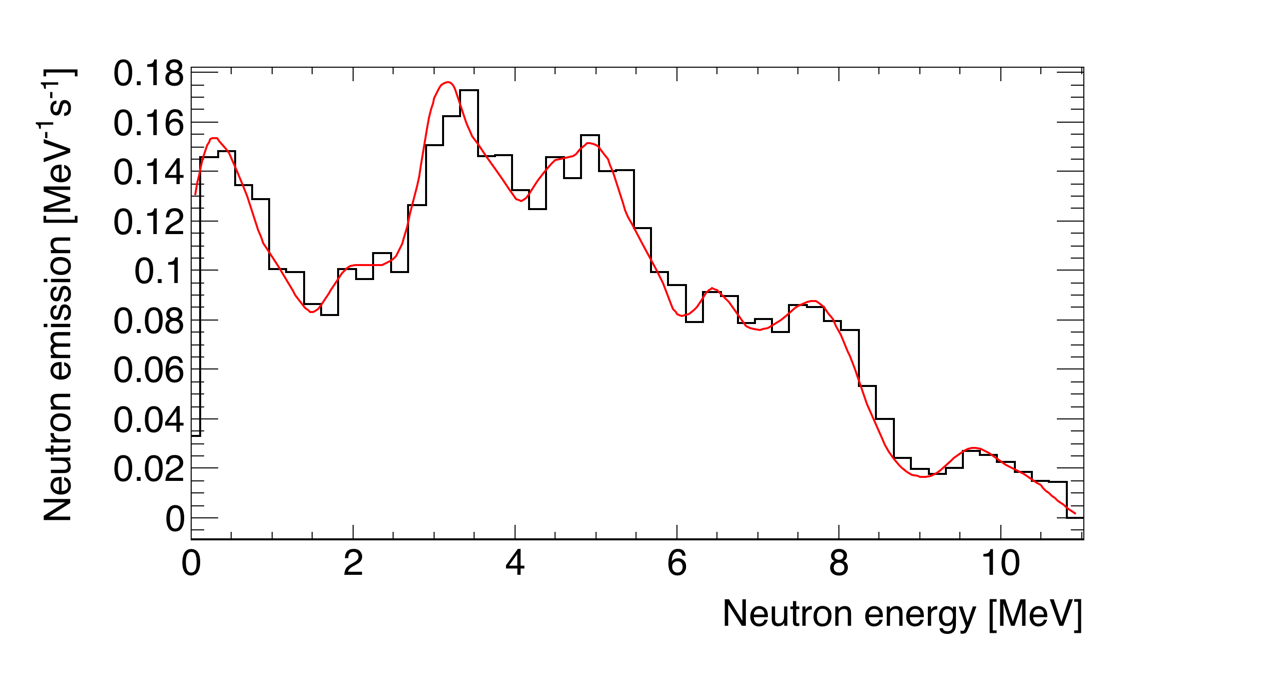}
\caption{The simulated ${}^{241}$Am-${}^{9}$Be source neutron emission spectrum (black) and the ISO 8529 \citep{iso} reference one (red).}
\label{ambe}
\end{figure}

The exact experimental configuration described above, was simulated by using GEANT4 \citep{agostinelli}. The two emission components of the ${}^{241}$Am-${}^{9}$Be source ie. the neutron emission and the gamma emission were simulated independently. In figure \ref{ambe} is shown the simulated neutron emission spectrum of the ${}^{241}$Am-${}^{9}$Be source compared with the reference emission spectrum ISO8529 \citep{iso} we used as input at the simulations.


The \enquote{\texttt{FTFP\_BERT\_HP\_LIV}} physics list was chosen in the simulation to include the high precision transport model \citep{geant4manual} for neutrons with energy lower than 20 MeV and also to include the "Livermore" model \citep{cirrone} for a more detailed description of the low energy electromagnetic interactions ($\sim$keV). 

\begin{figure}[!h]
\centering
\includegraphics[width=80mm]{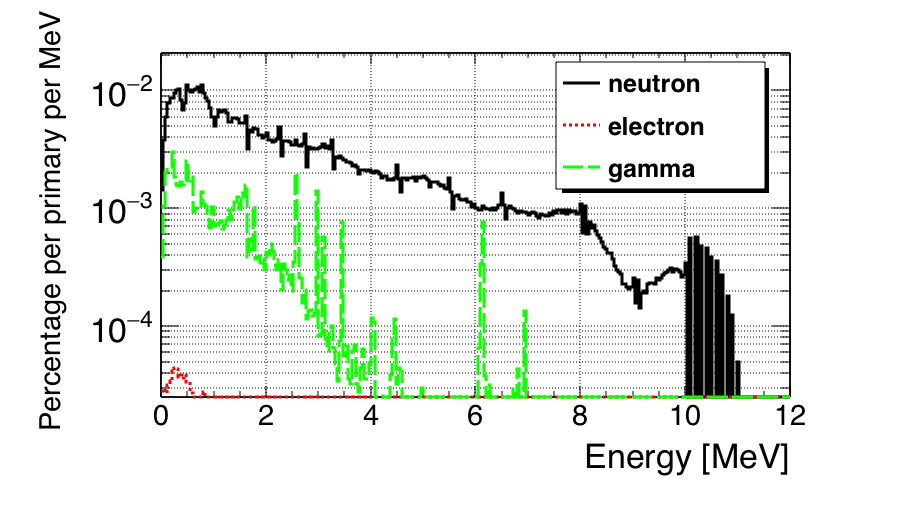}
\caption{The simulated number of electron, neutron and gamma particles that enter into the detector volume per primary neutron.}
\label{flux}
\end{figure}

First we are going to discuss the neutron emission simulation, ie. the energy deposition events registered in the gaseous volume of the detector are coming from neutron primaries and secondary particles which are produced by the interaction of neutrons with the materials of the experimental setup. In figure \ref{flux} we present the number of neutron, electron and gamma particles that enter into the gaseous volume of the detector during the simulation of the neutron emission of the ${}^{241}$Am-${}^{9}$Be source.

\begin{figure}[!h]
\centering
\includegraphics[width=90mm]{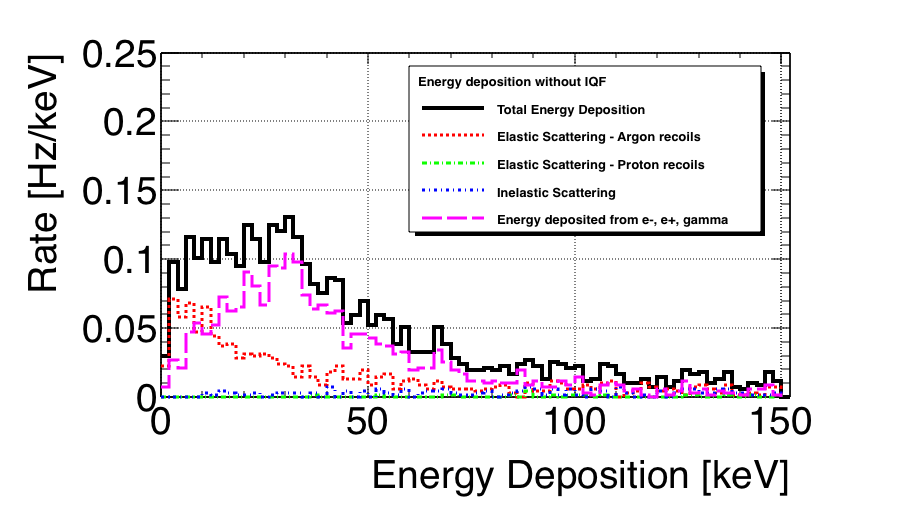}
\caption{The energy deposition spectra from the different energy depositing particles coming from the neutron emission of the ${}^{241}$Am-${}^{9}$Be source, without considering the ionization quenching factor.}
\label{res0}
\end{figure}

The energy deposited by these particles in the detector volume, taking into account all the different physical processes that they undergo is shown in figure \ref{res0}. The contribution from all particles forms the total energy deposition spectrum. The GEANT4 simulation output for the energy deposition of the low energy ions represents the total kinetic energy dissipated inside the medium but there is no differentiation between energy deposition leading to ionization of the medium and non-ionizing energy deposition as for example the ion-atom elastic scattering, which can be an important factor in the case of heavy ions. For this reason the ionization quenching factor should be taken into account.

\begin{figure}[!h]
\centering
\includegraphics[width=80mm]{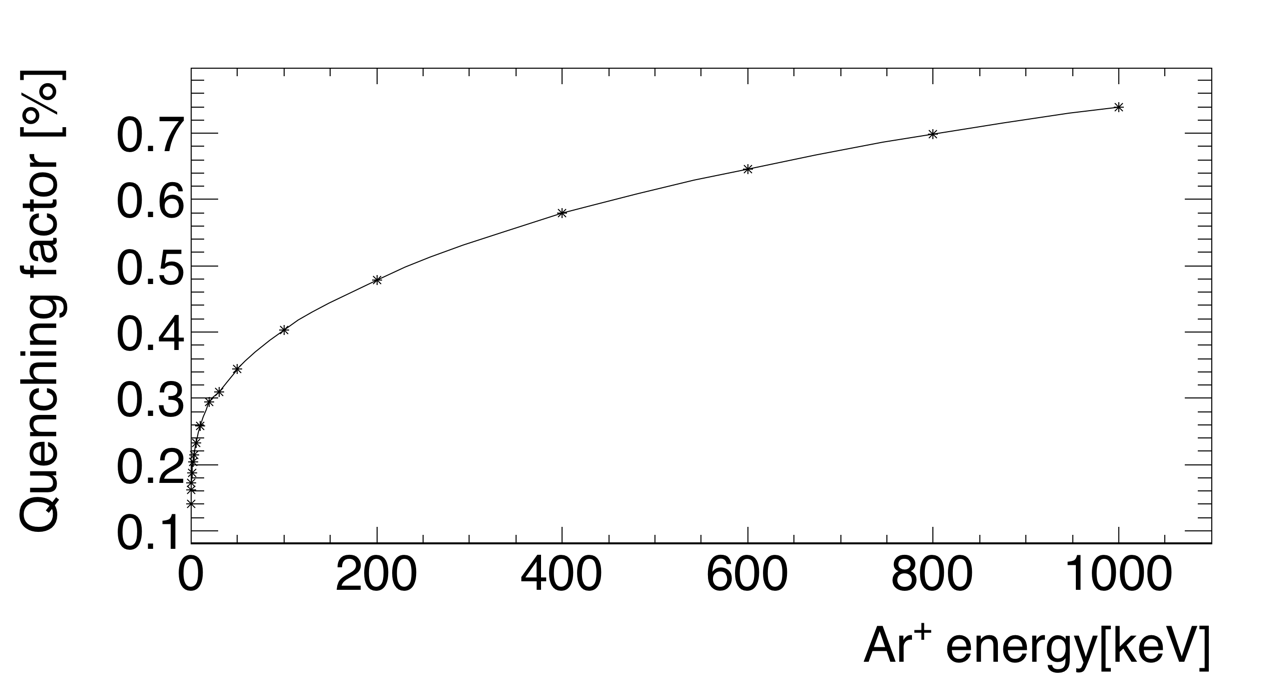}
\caption{The ionization quenching factor for Ar:CH${}_{4}\ (98:2)$ gas mixture calculated using SRIM.}
\label{iqf}
\end{figure}

\subsection{SRIM simulation of the Ionization Quenching Factor (IQF)} 

The ionization quenching factor (IQF) can be described as the fraction of kinetic energy of an ion that is dissipated in a medium in the form of ionization electrons and excitation of the atomic and quasi-molecular states \citep{platzman}. The other fraction of the kinetic energy of the ion is lost in non ionizing interactions like the ion-atom elastic scattering. This definition can be translated in the case of a proportional counter as the ratio of the ionization produced by a heavy ion with respect to the ionization produced by an electron of the same energy. Due to very few results published regarding measurements of the quenching factor in different materials and especially gases, we have to rely on software such as SRIM \citep{srim} to estimate the quenching factor for our gas mixture.  In SRIM the total energy loss of a heavy ion is divided between the "electronic" energy losses (which correspond to energy dissipated due to the interaction of an ion with the atomic electrons of a medium) and "nuclear" energy losses (which correspond to energy dissipated due to the elastic scattering of the ion with the atomic nucleus) \citep{ziegler2008srim}. For high energy ions the total energy loss is dominated by the electronic energy losses, but for ions with velocity lower than the Bohr orbital velocity of atomic electrons ($v_{ion} \leq v=e^{2}/\hbar=c/137$) \citep{evans} the nuclear energy losses become predominant. Using SRIM we estimated the IQF for the Ar:CH${}_{4}\ (98:2)$ gas mixture by calculating the ratio between the electronic energy losses over the kinetic energy of the Ar${}^{+}$ ion, the result is present at figure \ref{iqf}.

\begin{figure}[!h]
\centering
\includegraphics[width=90mm]{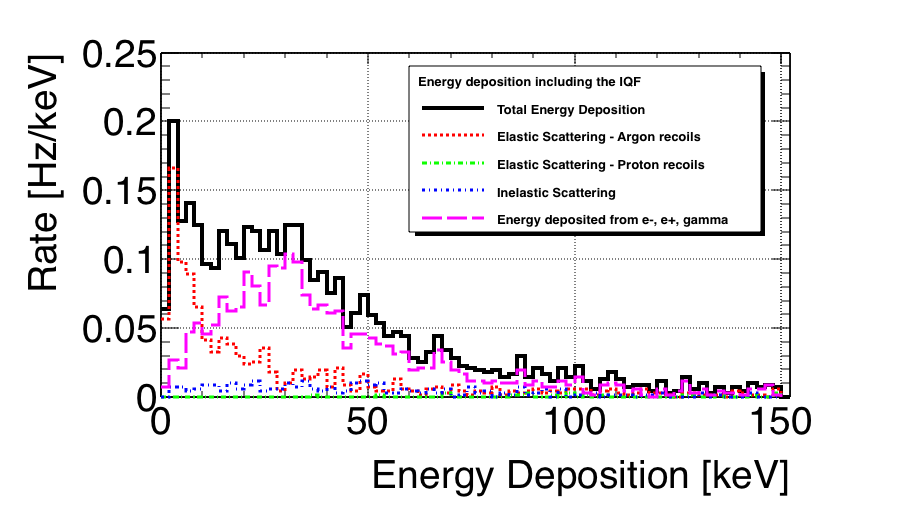}
\caption{The energy deposition spectra from the different energy depositing particles coming from the neutrons emitted by the ${}^{241}$Am-${}^{9}$Be source with the detector materials, taking into account the ionization quenching factor.}
\label{res1}
\end{figure}

The SRIM calculated IQF was used to estimate which part of the Ar recoil energy deposition spectrum, simulated using GEANT4, corresponds to energy deposition only due to electronic losses. In figure \ref{res1} we present the total energy deposition spectrum after the transformation of the Ar recoil energy deposition spectrum of figure \ref{res0} by the inclusion of the IQF. It is evident that because of the reduction of the accounted energy deposition by Ar recoils, the Ar recoil energy deposition distribution has shifted to lower energies.

\begin{figure}[!h]
\centering
\includegraphics[width=90mm]{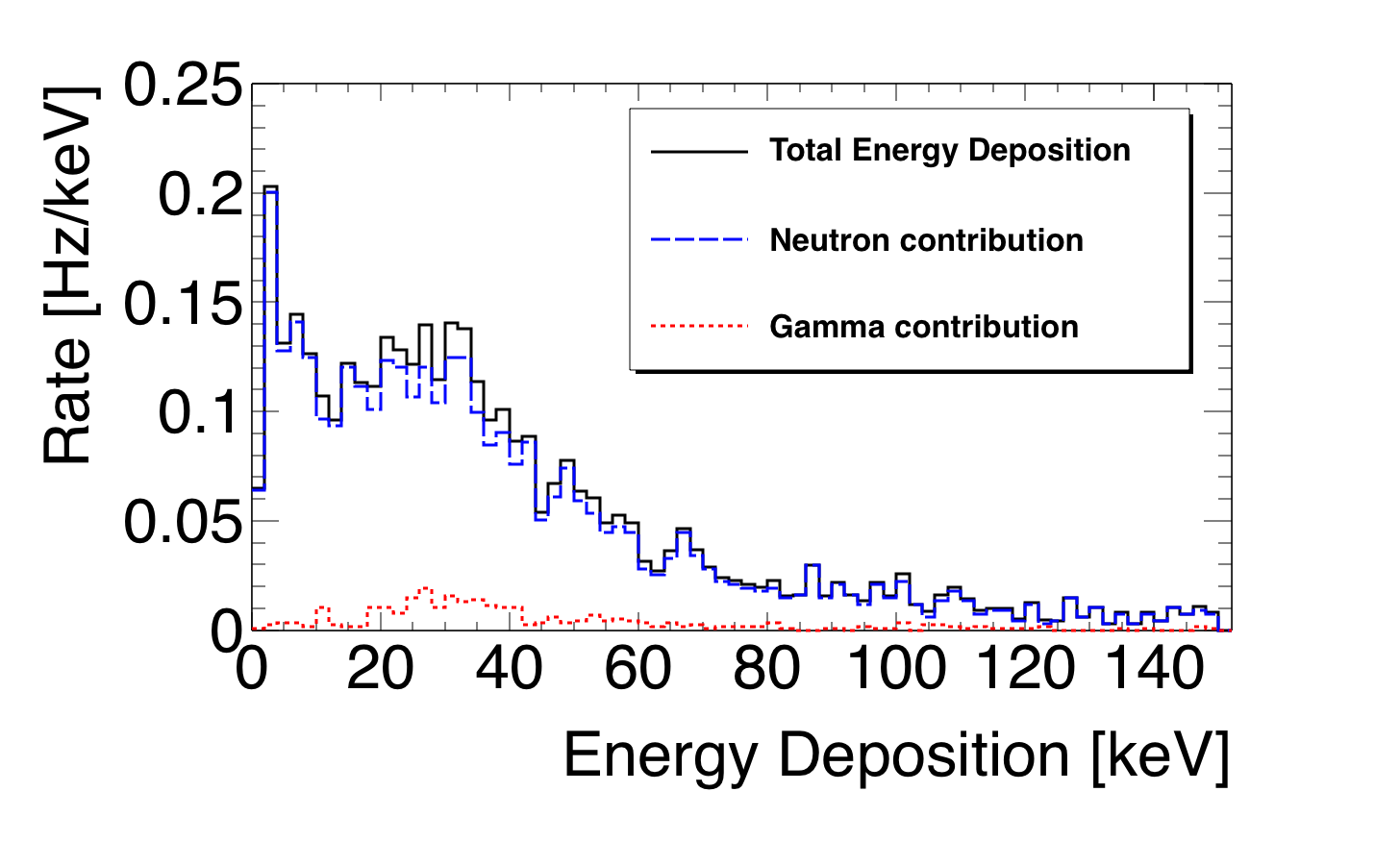}
\caption{The simulated total energy deposition (taking into account the IQF) from the neutron emission component and the gamma emission component of the ${}^{241}$Am-${}^{9}$Be source.}
\label{res2}
\end{figure}

To get the final energy deposition spectrum we have to include the contribution by events coming from the gamma emission component of the ${}^{241}$Am-${}^{9}$Be source. The energy deposition spectra of the neutron and gamma emission components of the source along with their sum are presented in figure \ref{res2}. 

\FloatBarrier
\subsection{Comparison between the measured and the simulated energy deposition spectra}

The first observation we can make by looking at figure \ref{edepcuts}-bottom and the simulation result (figure \ref{res2}) is that the energy range below 20 keV is dominated by the energy deposition of nuclear recoils while the "peak" around $\sim35$ keV is mainly due to the gamma rays emitted by the interaction of the source neutrons with the detector materials.

\begin{figure}[!h]
\centering
\includegraphics[width=90mm]{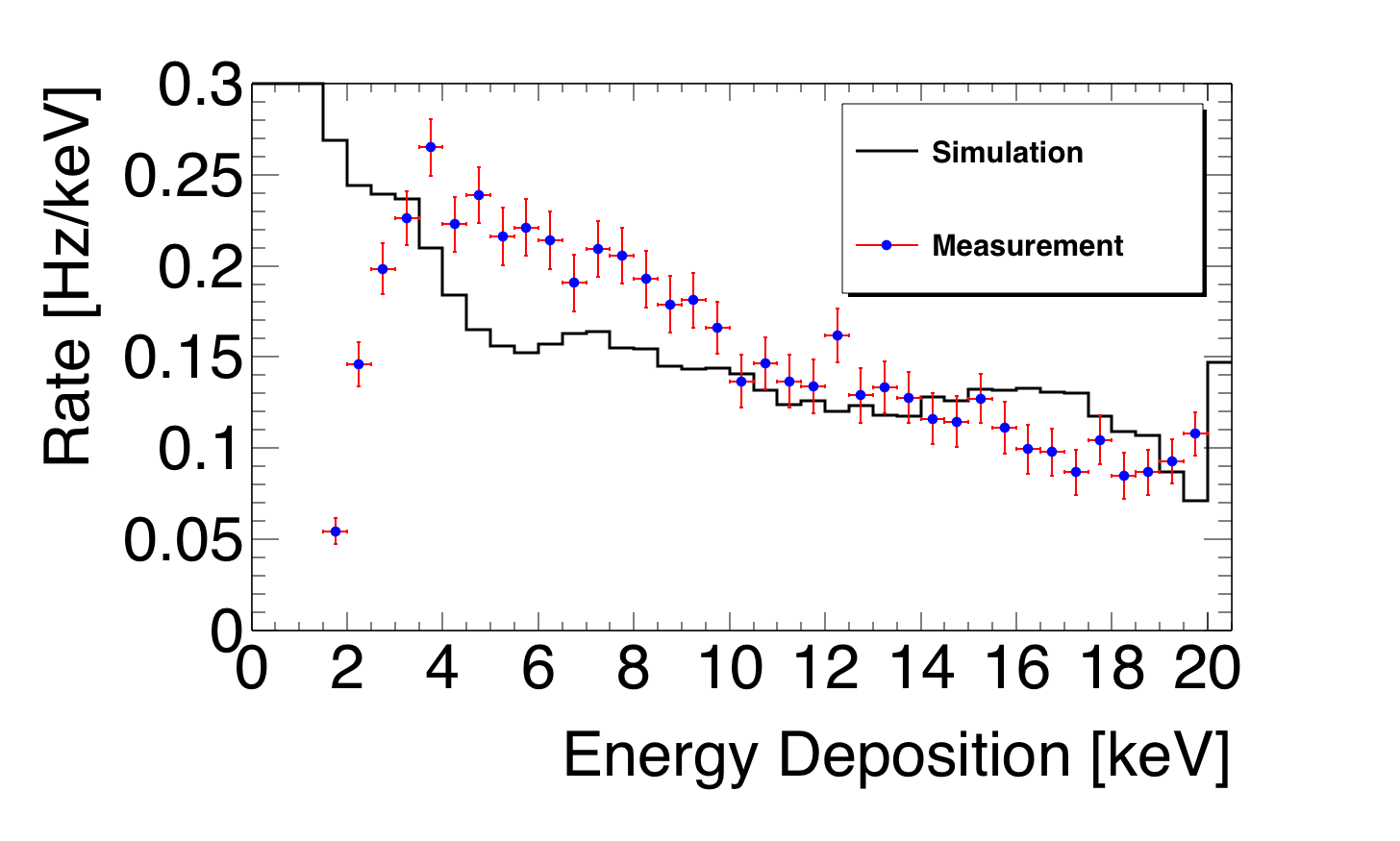}
\caption{Comparison between the measured spectrum and the simulated energy deposition spectrum up to 20 keV. The histogram represented by the black line corresponds to the simulated spectrum after the Gaussian smearing and the blue dots with the error bars correspond to the measured spectrum.}
\label{res3}
\end{figure}

It is interesting is to attempt a quantitative comparison between the measured and the simulated spectra below 20 keV, where the recoil contribution is predominant. For this reason the total energy deposition spectrum of figure \ref{res2} was smeared with a Gaussian distribution in order to take into account the detector response to ionizing radiation in terms of energy resolution. The Gaussian distribution standard deviation was chosen in a way so that we incorporate the energy dependence of the energy resolution following the relationship \citep{knoll}
\begin{equation}
\sigma(E) = \frac{\sigma_{0}}{\sqrt{E}}       
\end{equation}
where $\sigma(E)$ is the standard deviation of the Gaussian distribution of a given energy, $\sigma_{0}$ is the proportionality constant and $E$ is the energy. The proportionality constant was calculated from the energy resolution of the calibration measurements. The simulated energy deposition spectrum after the Gaussian smearing and the measured spectrum after proper normalization are presented in figure \ref{res3}. 
The acquisition threshold set at $\sim2$ keV affects the first part of the spectrum up $\sim3$ keV and does not allow direct comparison in this part. After this point we can see that the two spectra are in reasonable agreement. The differences we observe between the two spectra can be attributed to two main factors  a) the estimation of the IQF by SRIM and b) the approach used to simulate the response of the detector which can be considered as a first approximation. The comparison between the two spectra gives an indication that the IQF values predicted by SRIM are rather good but we cannot have an accurate validation. 

\FloatBarrier
\section{Conclusions-Prospects}
The SPC is capable of detecting very low energy nuclear recoils (in the few keV energy region), thus being a very good candidate for rare event searches that require a low energy threshold for nuclear recoils. 
The results of the simulations seem to be in reasonable agreement with the measurements. However, this agreement between the two can be improved by performing a series of actions, such as the experimental determination of the IQF in the energy range of interest and the comparison with the SRIM predictions, as well as the inclusion in the simulation of the physical processes that take place during the detector operation, resulting in a more detailed estimation of the detector response to ionizing radiation.

Along with these actions we intend to increase the pressure in which we operate up to 5 bar and try to detect heavy recoils at energies below 1 keV (ultra low energy recoil detection). We also plan like to perform background studies to improve the recoil to background events ratio.

\section*{Acknowledgments}

This research/publication has been co-financed by the European Union (European Social Fund – ESF) and Greek national funds through the Operational Program "Education and Lifelong Learning" of the National Strategic Reference Framework (NSRF) - Research Funding Program: “THALIS – HELLENIC OPEN UNIVERSITY- Development and Applications of Novel Instrumentation and Experimental Methods in Astroparticle Physics”.
\\
\\
This work is funded by the French National Research Agency (ANR-15-CE31-0008).
\\	 
\\
This research/publication was performed in the frame of the NEWS-G Collaboration.


\bibliography{mybibfile}

\end{document}